\documentclass[12pt]{article}

\usepackage[applemac]{inputenc}
\usepackage{amsmath,amssymb,fullpage}
\usepackage{showlabels}

\newtheorem{definition}{Definition}[section]
\newtheorem{example}{Example}[section]
\newtheorem{theorem}[definition]{Theorem}

\newtheorem{remark}[definition]{ \it Remark}
\newtheorem{coro}[definition]{Corollary}
\newtheorem{proposition}[definition]{Proposition}
\newtheorem{lemma}[definition]{Lemma}

\numberwithin{equation}{section}

\def\1B{\text{1\!\!I}}

\begin{document}
\title{A white noise approach to insider trading}

\author{
Bernt \O ksendal$^{1,2}$ 
\and Elin Engen R\o se$^{1}$
}

\date{20 August 2015}

\footnotetext[1]{Department of Mathematics, University of Oslo, P.O. Box 1053 Blindern, N--0316 Oslo, Norway \\
Emails: {\tt oksendal@math.uio.no},{\tt elinero@math.uio.no}} 
\footnotetext[2]{Norwegian School of Economics (NHH), Helleveien 30, N--5045 Bergen, Norway.}

\maketitle



\begin{abstract}

We present a new approach to the optimal portfolio problem for an insider with logarithmic utility.
Our method is based on white noise theory, stochastic forward integrals, Hida-Malliavin calculus and the Donsker delta function.

\end{abstract}

\section{Introduction}\label{intro}

The purpose of this paper is use concepts and methods from anticipating stochastic calculus, particularly from white noise theory and Hida-Malliavin calculus, to study optimal portfolio problems for an insider in a financial market driven by Brownian motion $B(t)$.
Our basic problem setup is related to the setup in \cite{PK}:

We assume that the insider at any time $t \in [0,T]$ has access to the information ($\sigma$-algebra) $\mathcal{F}_t$ generated by the driving Brownian motion up to time $t$, and in addition knows the value of some $\mathcal{F}_{T_0}$-measurable random variable $Y$, where $T_0 > T$ is some given future time. With this information flow $\mathbb{H} = \{ \mathcal{H}_t \}_{t \in [0,T]}$ with $\mathcal{H}_t = \mathcal{F}_t \vee \sigma(Y)$ to her disposal, she tries to find  the $\mathbb{H}$-adapted portfolio $\pi^{*}$ that maximises the expected logarithmic utility of the corresponding wealth at a given terminal time $T < T_0$. 

In \cite{PK} is it assumed that the insider filtration $\mathbb{H}$ allows an \emph{enlargement of filtration}, i.e. that there exists an $\mathbb{H}$-adapted process $\alpha(s)$  such that 
\begin{equation} \label{eq1.1}
\tilde{B}(t) := B(t) - \int_0^t \alpha(s) ds
\end{equation}
is a Brownian motion with respect to $\mathbb{H}$.
If this holds, the original problem, which was a priori a problem with anticipating stochastic calculus, can be transformed back to a semimartingale setting and (in some cases) solved using classical solution methods.
In terms of the process $\alpha$, \cite{PK} prove that the optimal insider portfolio can be written
\begin{equation} \label{eq1.2}
\pi^{*}(t) = \frac{b(t) - r(t)}{\sigma^2(t)} + \frac{\alpha(t)}{\sigma(t)}
\end{equation}
where $r(t)$ is the interest rate of the risk free asset, and $b(t), \sigma(t)$ are the drift term and the volatility of the risky asset, respectively.

In the present paper, we do not assume \eqref{eq1.1}, but in stead we follow the approach in \cite{BO} and in the recent paper \cite{DrO1} and work with \emph{anticipative stochastic calculus}. This means that the stochastic integrals involved in the anticipating insider portfolio are represented by \emph{forward integrals}. The forward integral, originally introduced in \cite{RV}, is an extension of the It\^ o integral, in the sense that it coincides with the It\^ o integral if the integrand is adapted. (See below.) It was first applied to insider trading in \cite{BO}, where it is pointed out why this integral appears naturally in the modelling of portfolio generated wealth processes in insider trading. In \cite{BO} a kind of converse to the result in \cite{PK}  is proved, namely that if an optimal insider portfolio exists, then the underlying Brownian motion $B(t)$ is indeed a semimartingale with respect to the insider filtration $\mathbb{H}$, and hence \eqref{eq1.1} holds.

The present paper differs also fundamentally from \cite{BO}, because we use white noise theory and Hida-Malliavin calculus to solve the anticipative optimal portfolio problem directly. The paper closest to ours is \cite{DrO1}. Indeed, our paper might be regarded as a discussion of a special case in \cite{DrO1}, although the method used in our paper is different and specially adapted to the logarithmic utility case. One of our main results is that if the \emph{Donsker delta function} $\delta_Y(y)$ of $Y$ exists in the Hida space $(\mathcal{S})^{*}$ of stochastic distributions, and the conditional expectations
$E[\delta_Y(y) | \mathcal{F}_t]$ and $E[D_t \delta_Y(y) | \mathcal{F}_t]$ both belong to $L^2(\lambda \times P)$, where $\lambda$ is Lebesgue measure on $[0,T]$ and $P$ is the probability law of $B(\cdot)$, then the optimal insider portfolio is
\begin{equation}\label{eq1.3}
\pi^{*}(t) = \frac{b(t)-r(t)}{\sigma^2(t)} + \frac{E[D_t \delta_Y(y)\mid \mathcal{F}_t]_{y=Y}}{\sigma(t)E[\delta_Y(y)\mid \mathcal{F}_t]_{y=Y}}
\end{equation}
where $D_t$ denotes the Hida-Malliavin derivative at $t$.
See Theorem 3.1.

Comparing \eqref{eq1.2} and \eqref{eq1.3} we get the following \emph{enlargement of filtration formula}, which is of independent interest (see Theorem 3.2):
\begin{equation}
\alpha(t) = \frac{E[D_ t \delta_Y(y)\mid \mathcal{F}_t]_{y=Y}}{E[\delta_Y(y)\mid \mathcal{F}_t]_{y=Y}}.
\end{equation}
For more general results in this direction, see \cite{DrO2}.\\

For simplicity we only discuss the Brownian motion case in this paper. For more information about Hida-Malliavin calculus in a white noise setting and extensions to
L\' evy processes and more general insider control problems, see \cite{DrO1}. 
\section{Background in white noise theory and Hida-Malliavin calculus}

In this section we summarise the basic notation and results we will need form white noise theory and the associated Hida-Malliavin calculus. For more details see e.g. \cite{BBS}, \cite {DOP}, \cite{DMOP1}, \cite{R}, \cite{DO}, \cite{DrO1} and the references therein. For a general introduction to white noise theory see \cite{HKPS}, \cite{HOUZ} and \cite{O}.
\\
\subsection{List of notation}
\vskip 0.5cm
\begin{itemize}
\item
$F \diamond G =$ the Wick product of random variables $F$ and $G$.
\item
$F^{ \diamond n} = F \diamond F \diamond F ... \diamond F$ (n times). (The $n$'th Wick power of $F$).
\item
$\exp^{\diamond}(F) = \Sigma_{n=0}^{\infty} \frac{1}{n!} F^{\diamond n}$ (The Wick exponential of $F$.)
\item
$\varphi^{\diamond}(F)$ = the Wick version of the random variable $\varphi(F)$
\item
$D_t F =$ the Hida-Malliavin derivative of $F$ at $t$ with respect to $B(\cdot)$.\\
This is denoted by $\partial_t F$ in \cite{O}, see page 30 there.
\item
$D_{t}(\varphi( F)) = ((\varphi)')(F) \diamond D_{t} F$. (The Hida-Malliavin chain rule.)
\item
$D_{t}(\varphi^{\diamond}( F)) = ((\varphi)')^{\diamond}(F) \diamond D_{t} F$. (The Wick chain rule.)
\item
$(\mathcal{S}), (\mathcal{S})^{'} $ = the Hida stochastic test function space and stochastic distribution space, respectively.
\item
$(\mathcal{S}) \subset L^2(P) \subset (\mathcal{S})^{'} $.\\
Here, as usual, $L^2(P)$ is the set of random variables $F$ with $E[F^2] < \infty$ , where $E[\cdot]$ denotes expectation with respect to the probability measure $P$.
\end{itemize}
For more information about the Wick calculus we refer to Section 7 of Chapter 1 in \cite{O}.
\subsection{ The forward integral with respect to Brownian motion}

The forward integral with respect to Brownian motion was first defined in
the seminal paper \cite{RV} and further studied in \cite{RV1}, \cite{RV2}. This integral was
introduced in the modelling of insider trading in \cite{BO} and then applied by several authors
in questions related to insider trading and stochastic control with
advanced information (see, e.g., \cite{DMOP2}). The forward integral was later extended to Poisson random measure integrals in \cite{DMOP1}.

\begin{definition}
We say that a stochastic process $\phi = \phi(t), t\in[0, T ]$, is
\emph{forward integrable} (in the weak sense) over the interval $[0, T ]$ with respect to
$B$ if there exists a process $I = I(t), t\in[0, T ]$, such that
\begin{equation}
\sup_{t\in[0,T ]}|\int_0^t\phi(s)\frac{B(s+\epsilon)-B(s)}{\epsilon}ds-I(t)|\rightarrow 0, \quad \epsilon\rightarrow0^+
\end{equation}
in probability. In this case we write
\begin{equation}
I(t) :=\int_0^t\phi(s)d^-B(s), t\in[0, T ],
\end{equation}
and call $I(t)$ the \emph{forward integral} of $\phi$ with respect to $B$ on $[0, t]$.
\end{definition}

The following results give a more intuitive interpretation of the forward integral
as a limit of Riemann sums:
\begin{lemma}
Suppose $\phi$ is c\`{a}gl\`{a}d and forward integrable. Then
\begin{equation}
\int_0^T\phi(s)d^-B(s) = \lim _{\triangle t\rightarrow0}\sum_{j=1}^{J_n}\phi(t_{j-1})(B(t_j)-B(t_{j-1}))
\end{equation}
with convergence in probability. Here the limit is taken over the partitions \\
$0 =t_0 < t_1 < ... < t_{J_n}= T$ of $t\in[0, T ]$ with $\triangle t:= \max_{j=1,...,J_n}(t_j- t_{j-1})\rightarrow 0,
n\rightarrow\infty.$
\end{lemma}

\begin{remark}

From the previous lemma we can see that, if the integrand $\phi$
is $\mathcal{F}$-adapted, then the Riemann sums are also an approximation to the It\^{o}
integral of $\phi$ with respect to the Brownian motion. Hence in this case the
forward integral and the It\^{o} integral coincide. In this sense we can regard
the forward integral as an extension of the It\^{o} integral to a nonanticipating
setting.
\end{remark}

We now give some useful properties of the forward integral. The
following result is an immediate consequence of the definition.

\begin{lemma}
 Suppose $\phi$ is a forward integrable stochastic process and $G$ a
random variable. Then the product $G\phi$ is forward integrable stochastic process and
\begin{equation}
\int_0^TG\phi(t)d^-B(t) = G\int_0^T\phi(t)d^-B(t)
\end{equation}
\end{lemma}

The next result shows that the forward integral is an extension of the
integral with respect to a semimartingale:
\begin{lemma}
Let $\mathbb{G} := \{\mathcal{G}_t, t\in[0, T ]\} (T > 0)$ be a given filtration. Suppose
that
\begin{enumerate}
\item $B$ is a semimartingale with respect to the filtration $\mathbb{G}$.
\item $\phi$ is $\mathbb{G}$-predictable and the integral
\begin{equation}
\int_0^T\phi(t)dB(t),
\end{equation}
with respect to $B$, exists.\\
Then $\phi$ is forward integrable and
\begin{equation}
\int_0^T\phi(t)d^-B(t)=\int_0^T\phi(t)dB(t),
\end{equation}
\end{enumerate}
\end{lemma}

We now turn to the It\^{o} formula for forward integrals. In this connection it is
convenient to introduce a notation that is analogous to the classical notation
for It\^{o} processes.
\begin{definition}
A \emph{forward process} (with respect to $B$) is a stochastic process
of the form
\begin{equation}\label{forward form 1}
X(t) = x +\int_0^tu(s)ds +\int_0^tv(s)d^-B(s), \quad t\in[0, T ],
\end{equation}
($x$ constant), where $\int_0^T|u(s)|ds <\infty, \mathbf{P}$-a.s.
and $v$ is a forward integrable stochastic process. A shorthand notation for
(\ref{forward form 1}) is that
\begin{equation}
d^-X(t) = u(t)dt + v(t)d^-B(t).
\end{equation}
\end{definition}

\begin{theorem}{\emph{The one-dimensional It\^{o} formula for forward integrals.}} \\
Let
\begin{equation}
d^-X(t) = u(t)dt + v(t)d^-B(t)
\end{equation}
be a forward process. Let $f\in\mathbf{C}^{1,2}([0, T ]\times\mathbb{R})$ and define
\begin{equation}
Y (t) = f(t,X(t)), \quad t\in[0, T ].
\end{equation}
Then $Y (t), t\in[0, T ]$, is also a forward process and
\begin{equation}
d^-Y (t) = \frac{\partial f}{\partial t}(t,X(t))dt+\frac{\partial f}{\partial x}(t,X(t))d^-X(t)+\frac{1}{2}\frac{\partial^2f}{\partial x^2} (t,X(t))v^2(t)dt.
\end{equation}
\end{theorem}

We also need the following forward integral result, which is obtained by an adaptation of the proof of Theorem 8.18 in \cite{DOP}:
\begin{proposition}

Let $\varphi$ be a c\`{a}gl\`{a}d and forward integrable process in $L^2(\lambda \times P)$. Then
$$E[D_{s^{+}} \varphi(s)|\mathcal{F}_s]:= \lim _{\epsilon \rightarrow 0^{+}} \frac{1}{\epsilon}\int_{s-\epsilon}^s E[D_s \varphi(t)|\mathcal{F}_s]dt$$
exists in $L^2(\lambda \times P)$ and 
\begin{align}\label{eq2.12}
& E[\int_0^T \varphi(s) d^{-}B(s)] = E[\int_0^T E[D_{s^{+}} \varphi(s)|\mathcal{F}_s] ds].
\end{align}\\
\end{proposition}

Similar definitions and results can be obtained in the Poisson random measure case. See \cite{DMOP1} and \cite{DOP}.

\subsection{The Donsker delta function}

As in \cite{P}, Chapter VI, we define the \emph{regular conditional distribution} with respect to $\mathcal{F}_t$ of a given real random variable $Y$, denoted by $Q_t(dy)=Q_t(\omega,dy)$, by the following properties:
\begin{itemize}
\item
For any Borel set $\Lambda \subseteq \mathbb{R}, Q_t(\cdot, \Lambda)$ is a version of $E[\chi_{Y \in dy} | \mathcal{F}_t]$
\item
For each fixed $\omega, Q_t(\omega),dy)$ is a probability measure on the Borel subsets of $\mathbb{R}$
\end{itemize}

It is well-known that such a regular conditional distribution always exists. See e. g. \cite{B}, page 79.

From the required properties of $Q_t(\omega,dy)$ we get the following formula
\begin{equation}
\int_{\mathbb{R}} f(y) Q_t(\omega,dy) = E[ f(Y) | \mathcal{F}_t]
\end{equation}
Comparing with the definition of the Donsker delta function, we obtain the following representation of the regular conditional distribution:

\begin{proposition}
Suppose $Q_t(\omega,dy)$ is absolutely continuous with respect to Lebesgue measure on $\mathbb{R}$. Then
\begin{equation}
\frac{Q_t(\omega,dy)}{dy} = E[ \delta_Y(y) | \mathcal{F}_t] 
\end{equation}
\end{proposition}
Explicit formulas for the Donsker delta function are known in many cases. For the Gaussian case, see Section 3.2. For details and more general cases, see \cite{AaOU}, \cite{DO},\cite{DiO},\cite{MOP}, \cite{MP}, \cite{LP} and \cite{DrO1}. See also Example 22 in Chapter 1 in \cite{O}.

\section{The market model and the optimal portfolio problem for the insider}\label{sec3}
Suppose we have a market with the following two investment possibilities:
\begin{itemize}
\item
A risk free investment (e.g. a bond or a (safe) bank account), whose unit price $S_0(t)$ at time $t$ is described by
\begin{equation}
\begin{cases}
dS_0(t) = r(t)S_0(t) dt; \quad 0 \leq t \leq T\\
S_0(0) = 1
\end{cases}
\end{equation}
\item
A risky investment, whose unit price $S(t)$ at time $t$ is described by a stochastic differential equation (SDE) of the form
\begin{equation}
\begin{cases}
dS(t) = S(t)[b(t) dt + \sigma(t) dB(t)] ; \quad 0 \leq t \leq T\\
S(0) > 0.
\end{cases}
\end{equation}\\
\end{itemize}

Here $T$ is a fixed, given constant terminal time, $r(t)=r(t,\omega)$, $b(t) =b(t,\omega)$ and  $\sigma(t)=\sigma(t,\omega)$ are given $\mathbb{F}$-adapted processes, and $B(t)$ is a Brownian motion on a filtered probability space $(\Omega, \mathbb{F}=\{\mathcal{F}_t\}_{t\geq 0},P)$. We assume that
$\sigma(t)>0$ is bounded away from 0, and that 
\begin{equation}
E[\int_0^T \{|b(t)| + |r(t]| + \sigma^2(t) \} dt] < \infty.
\end{equation}

\subsection{The optimal portfolio problem}

We consider an optimal portfolio problem for a trader with inside information. Thus we assume that a filtration $\mathbb{H} = \{{\mathcal{H}_t}\}_{t \geq 0}$ is given, which is an \emph{insider filtration}, in the sense that 
$$\mathcal{F}_t \subseteq \mathcal{H}_t$$ 
for all $t$.

Suppose a trader in this market has the inside information represented by $\mathbb{H}$ to her disposal. Thus at any time $t$ she is free to choose the \emph{fraction} $\pi(t)$ of her current portfolio wealth $X(t)=X^{\pi}(t)$ to be invested in the risky asset, and this fraction is allowed to depend on $\mathcal{H}_t$, not just $\mathcal{F}_t$. If the portfolio is \emph{self-financing} (which we assume), the the corresponding wealth process
 $X(t)=X^{\pi}(t)$ will satisfy the SDE 
\begin{equation} \label{eq3.3}
dX(t) = (1-\pi(t))X(t) \rho(t)(t) dt + \pi(t) X(t^-)[\mu(t) dt + \sigma(t) dB^-(t)]. 
\end{equation}
For simplicity we put $X(0) = 1.$
Since we do not assume that $\pi$ is $\mathbb{F}$-adapted, the stochastic integrals in \eqref{eq3.3}  are \emph{anticipating}. Following the argument in \cite{BO} we choose to interpret the stochastic integrals as \emph{forward integrals}, indicated by $dB^-(t)$.

By the It\^ o formula for forward integrals the solution of this SDE \eqref{eq3.3} is
\begin{align}\label{eq3.4}
&X(t)= 
\exp[\int_{0}^{t} \{\rho(s) +[\mu(s)-\rho(s)]\pi(s)-\frac{1}{2}\sigma^{2}(s)\pi^{2}(s)\}ds
+ \int_{0}^{t}\pi(s)\sigma(s) d^-B(s)
]
\end{align}

Let $U: [0,\infty) \mapsto [- \infty, \infty)$ be a given \emph{utility function}, i.e. a concave function on $[0,\infty)$, smooth on $(0,\infty)$, and let $\mathcal{A}_{\mathbb{H}}$ be a given family of $\mathbb{H}$-adapted portfolios. The \emph{insider optimal portfolio problem} we consider, is the following:

\vskip 0.5cm

$\mathbb{PROBLEM}$ \text{    }
Find $\pi^* \in \mathcal{A}_{\mathcal{H}}$ such that
\begin{equation}\label{eq3.5}
sup_{\pi \in \mathcal{A}_{\mathcal{H}}} E[U(X_{\pi}(T))] = E[U(X_{\pi^*}(T))].
\end{equation}

In this paper we will restrict ourselves to consider the \emph{logarithmic utility} $U_0$, defined by
\begin{equation}\label{eq3.6}
U(x) =U_0(x) := \ln(x).
\end{equation}

We will also assume that the inside filtration is of \emph{initial enlargement} type, i.e.
\begin{equation}
\mathcal{H}_t = \mathcal{F}_t \vee Y
\end{equation}
for all $t$, where $Y$ is a given $\mathcal{F}_{T_0}$ -measurable random variable, for some $T_0 > T$.

Thus we assume that the trader at any time $t$ knows all the value of a given $\mathcal{F}_{T_0}$ -measurable random variable $Y$, together with the values of the underlying noise process $B(s)$ for all $s \leq t$. Thus $\pi(t)$ is assumed to be measurable with respect the $\sigma$-algebra $\mathcal{H}_t$ generated by $Y$ and $B(s)$ for all $s \leq t$. In particular, the trader knows at time $t$ the exact values of all the coefficients of the system and the values of the price processes at time $t$.\\

From \eqref{eq3.4} and \eqref{eq2.12} we get
\begin{align} \label{eq4.2}
&E[\ln(X^{\pi}(T)] = E[ \int_{0}^{T} \{r(s) +[b(s)-r(s)]\pi(s)-\frac{1}{2}\sigma^{2}(s)\pi^{2}(s)
\}ds + \int_{0}^{T}\pi(s)\sigma(s) d^-B(s)]\nonumber\\
&=E[ \int_{0}^{T} \{r(s) +[b(s)-r(s)]\pi(s)-\frac{1}{2}\sigma^{2}(s)\pi^{2}(s)+ \sigma(s) D_s \pi(s)\}ds]\nonumber\\
&=E[ \int_{0}^{T} E[r(s) +[b(s)-r(s)]\pi(s)-\frac{1}{2}\sigma^{2}(s)\pi^{2}(s) +  \sigma(s) D_s \pi(s)\mid \mathcal{F}_s] ds]
\end{align}
Here and in the following we use the notation
$$ D_s \pi(s) := D_{s^{+}} \pi(s) := \lim_{t \rightarrow s^{+}} D_t \pi(s)$$
where as before $D_t$ denotes the Hida-Malliavin derivative at $t$.
Since $\pi$ is assumed to be $\mathbb{H}$-adapted, it has the form
\begin{equation}\label{eq4.3}
\pi(t,\omega)= f(t,Y,\omega)
\end{equation}
for some function $f: [0,T] \times \mathbb{R} \times \Omega \to \mathbb{R}$ such that $f(\cdot,y)$ is $\mathbb{F}$-adapted for each $y \in \mathbb{R}$.

Thus we can maximize \eqref{eq4.2} over all $\pi \in \mathcal{A}_{\mathbb{H}} $ by maximizing over all functions $f(t,Y)$ the integrand 
\begin{equation} \label{eq4.4}
J(f):= E[(b(s)-r(s))f(s,Y)-\frac{1}{2}\sigma^{2}(s)f^{2}(s,Y) +  \sigma(s) D_s f(s,Y)\mid \mathcal{F}_s] 
\end{equation}
for each s. To this end, suppose the random variable $Y$ has a Hida-Malliavin differentiable Donsker delta function $\delta_Y(y) \in (\mathcal{S})^{'}$, so that
$$g(Y)=\int_{\mathbb{R}} g(y) \delta_Y(y) dy $$
for all functions $g$ such that the integral converges.
Then
\begin{equation}\label{eq4.5}
f(s,Y)=\int_{\mathbb{R}} f(s,y) \delta_Y(y) dy 
\end{equation},

\begin{equation}\label{eq4.6}
f^2(s,Y)=\int_{\mathbb{R}} f^2(s,y) \delta_Y(y) dy 
\end{equation}

and
\begin{equation}\label{eq4.7}
D_s f(s,Y) = \int_{\mathbb{R}} f(s,y) D_s \delta_Y(y) dy 
\end{equation}
Substituting this into \eqref{eq4.4} we get
\begin{align} \label{eq4.8}
&J(f):= E[\int_{\mathbb{R}} \{(b(s)-r(s))f(s,y) \delta_Y(y)-\frac{1}{2}\sigma^{2}(s)f^{2}(s,Y) \delta_Y(y)+  \sigma(s) f(s,y) D_s \delta_Y(y)\} dy\mid \mathcal{F}_s]\nonumber\\
&= \int_{\mathbb{R}} \{(b(s)-r(s))f(s,y) E[\delta_Y(y)\mid \mathcal{F}_s] -\frac{1}{2}\sigma^{2}(s)f^{2}(s,Y) E[\delta_Y(y)\mid \mathcal{F}_s] \nonumber\\
&+  \sigma(s) f(s,y) E[D_s \delta_Y(y)\mid \mathcal{F}_s]\} dy\mid \mathcal{F}_s]
\end{align}
We can maximize this over $f(s,y)$ for each $s,y$. If we assume that
\begin{equation}\label{eq4.9}
0 < E[\delta_Y(y)\mid \mathcal{F}_s] \in L^2(\lambda \times P) \text{ and }
E[D_s \delta_Y(y)\mid \mathcal{F}_s] \in L^2(\lambda \times P)
\end{equation}
for all $s,y$,
then we see that the unique maximizing value of $f(s,y)$ is
\begin{equation}\label{eq4.10}
f^{*}(s,y) = \frac{b(s)-r(s)}{\sigma^2(s)} + \frac{E[D_s \delta_Y(y)\mid \mathcal{F}_s]}{\sigma(t)E[\delta_Y(y)\mid \mathcal{F}_s]}
\end{equation}

We have proved the following, which extends a result in \cite{PK} (and is a special case of results in \cite{DrO1}):
\begin{theorem}{[Optimal insider portfolio]}

Suppose $Y$ has a Donsker delta function $\delta_Y(y)$ satisfying \eqref{eq4.9}. Then the optimal insider portfolio is given by
\begin{equation}\label{eq4.11}
\pi^{*}(s) = \frac{b(s)-r(s)}{\sigma^2(s)} + \frac{E[D_s \delta_Y(y)\mid \mathcal{F}_s]_{y=Y}}{\sigma(t)E[\delta_Y(y)\mid \mathcal{F}_s]_{y=Y}}
\end{equation}
\end{theorem}

Combining this result with the results of \cite{PK} and \cite{BO} given in the Introduction, we get the following result of independent interest. It is a special case of results in \cite{DrO2}: 

\begin{theorem}{[Enlargement of filtration and semimartingale decomposition]}

Suppose $Y$ has a Donsker delta function $\delta_Y(y)$ satisfying \eqref{eq4.9}.
Then the $\mathbb{F}-$Brownian motion $B$ is a semimartingale with respect to the inside filtration $\mathbb{H}$, and it has the semimartingale decomposition
\begin{equation}\label{eq4.12a}
B(t) = \tilde{B}(t) +\int_0^t \alpha(s) ds,
\end{equation}
where $\tilde{B}(s)$ is an $\mathbb{H}-$ Brownian motion, and $\alpha(s)$ (called the information drift) is given by
\begin{equation}\label{eq4.13a}
\alpha(s)=\frac{E[D_s \delta_Y(y)\mid \mathcal{F}_s]_{y=Y}}{\sigma(t)E[\delta_Y(y)\mid \mathcal{F}_s]_{y=Y}}.
\end{equation}

\end{theorem}
This result is a special case of semimartingale decomposition results for L\' evy processes in \cite{DrO2}. For information about enlargement of filtration in general, see \cite{JY} and \cite{J} and the references therein. 

\subsection{Examples}
\begin{example}
Consider the special case when $Y$ is a Gaussian random variable of the form
 \begin{equation} \label{eq4.12}
Y=Y(T_0), \text{  where  } Y(t) = \int_0^t \psi(s) dB(s); \text{ for } t \in [0,T_0]
\end{equation}
for some deterministic function $\psi \in L^2[0,T_0]$ with 
$$ \|\psi \|_{[t,T]}^2 := \int_t^T \psi(s)^2 ds > 0 \text{   for all   } t \in [0,T]$$

In this case it is well known that the Donsker delta function exists in $(\mathcal{S})^{'}$ and is given by
\begin{equation}\label{eq4.13}
\delta_Y(y) = (2\pi v)^{-\frac{1}{2}}
 \exp^{\diamond}[-\frac{(Y-y)^{2\diamond}}{2v}]
\end{equation}
where we have put $ v:= \|\psi \|_{[0,T_0]}^2$.
See e.g. \cite{AaOU}, Proposition 3.2.\\
Using the Wick rule when taking conditional expectation, using the martingale property of the process $Y(t)$ and applying Lemma 3.7 in \cite{AaOU} we get
\begin{align}\label{eq4.14}
E[\delta_Y(y) | \mathcal{F}_t] &= (2\pi v)^{-\frac{1}{2}} \exp^{\diamond}
[-E[\frac{(Y(T_0)-y)^{2\diamond}}{2v}  | \mathcal{F}_t]]\nonumber\\
&=(2\pi \|\psi \|_{[0,T_0]}^2)^{-\frac{1}{2}} \exp^{\diamond}[-\frac{(Y(t)-y)^{2\diamond}}{2\|\psi \|_{[0,T_0]}^2}]\nonumber\\
&=(2\pi \| \psi \|_{[t,T_0]}^2)^{-\frac{1}{2}} \exp[-\frac{(Y(t)-y)^{2}}{2 \| \psi \|_{[t,T_0]}^2}]
\end{align}
Similarly, by the Wick chain rule and Lemma 3.8 in \cite{AaOU} we get
\begin{align}\label{eq4.15}
E[D_t\delta_Y(y) | \mathcal{F}_t] & = - E[(2\pi v)^{-\frac{1}{2}}
 \exp^{\diamond}[-\frac{(Y(T_0)-y)^{2\diamond}}{2v}] \diamond \frac{Y(T_0)-y}{v} \psi(t) | \mathcal{F}_t]\nonumber\\
& = - (2\pi v)^{-\frac{1}{2}} \exp^{\diamond}[-\frac{(Y(t)-y)^{2\diamond}}{2v}] \diamond \frac{Y(t)-y}{v} \psi(t)\nonumber\\
& = - (2\pi \| \psi \|_{[t,T_0]}^2)^{-\frac{1}{2}} \exp [-\frac{(Y(t)-y)^2}{2\| \psi \|_{[t,T_0]}^2}]  \frac{Y(t)-y}{\| \psi \|_{[t,T_0]}^2} \psi(t)
\end{align}

Substituting \eqref{eq4.14} and \eqref{eq4.15} in \eqref{eq4.11} we obtain:

\begin{coro}
Suppose that Y is Gaussian of the form \eqref{eq4.12}. Then the optimal insider portfolio is given by
\begin{equation}
\pi^{*}(t) = \frac{b(t)-r(t)}{\sigma^2(t)} + \frac{(Y(T_0)-Y(t)) \psi(t)}{\sigma(t)\| \psi \|_{[t,T_0]}^2}
\end{equation}
\end{coro}
\end{example}

In particular, if $Y = B(T_0)$ we get the following result, which was first proved in \cite{PK} (by a different method):
\begin{coro}
Suppose that $Y= B(T_0)$. Then the optimal insider portfolio is given by
\begin{equation}
\pi^{*}(t) = \frac{b(t)-r(t)}{\sigma^2(t)} + \frac{B(T_0)-B(t)}{\sigma(t)(T_0-t)}
\end{equation}
\end{coro}

\end{document}